\documentclass[12pt,preprint]{aastex}

\cpright{AAS}{2010}
\shorttitle{Coronal Dimmings and Heat Waves}
\shortauthors{E. Robbrecht & Y.-M. Wang}

\begin{document}

\title{The Temperature-Dependent Nature of Coronal Dimmings}

\author{Eva Robbrecht\altaffilmark{1} and Yi-Ming Wang \altaffilmark{2}}
\altaffiltext{1}{Royal Observatory of Belgium, Ringlaan 3 - 1180 Brussels, Belgium} 
\altaffiltext{2}{Naval Research Laboratory, 4555 Overlook Ave SW, Washington, DC 20375-5352, USA}

\email{Eva.Robbrecht@oma.be, Yi.Wang@nrl.navy.mil}


\begin{abstract}
The opening-up of the magnetic field during solar eruptive events 
is often accompanied by a dimming of the local coronal emission.  From observations of 
filament eruptions recorded with the Extreme-Ultraviolet Imager on 
{\it STEREO} during 2008--2009, it is evident that these dimmings are 
much more pronounced in 19.5~nm than in the lower-temperature 
line 17.1~nm, as viewed either on the disk or above the limb.  
We conclude that most of the cooler coronal plasma is not ejected but remains 
gravitationally bound when the loops open up.  This result is consistent 
with Doppler measurements by Imada and coworkers, who found that the 
upflow speeds in a transient coronal hole increased dramatically 
above a temperature of 1 MK; it is also consistent with the quasistatic 
behavior of polar plumes, as compared with the hotter interplume regions 
that are the main source of the fast solar wind.  When the open flux 
reconnects and closes down again, the trapped plasma is initially heated 
to such high temperatures that it is no longer visible at  \ion{Fe}{9} 17.1~nm.  Correspondingly, 
17.1~nm images show a dark ribbon or ``heat wave'' propagating away 
from the polarity inversion line and coinciding with the 
brightened \ion{Fe}{15} 28.4~nm and \ion{Fe}{12} 19.5~nm post-eruptive loops and their footpoint areas.  
Such dark ribbons provide a clear example of dimmings that are not caused by a density depletion. 
The propagation of the ``heat wave'' is driven by the closing-down, 
not the opening-up, of flux and can be observed both off-limb and on-disk.
\end{abstract}



 

\section{Introduction}
Filaments eruptions and other ejections of mass from the Sun 
are often accompanied by a dimming of the local coronal emission 
at many different wavelengths, and by the formation of transient 
coronal holes (see, e.g., Harrison \& Lyons 2000; Kahler \& Hudson 2001; 
Harrison et al. 2003; Howard \& Harrison 2004; Attrill et al. 2006; 
Harra et al. 2007; Imada et al. 2007; Reinard \& Biesecker 2008, 2009; 
Jin et al. 2009; Dai et al. 2010).  The dimmings are in most cases 
caused by a decrease in the coronal density due to the opening-up of the 
magnetic field and escape of the entrained material into the heliosphere.  
The closing-down of the flux proceeds from the inside outward, with the 
field lines rooted nearest the photospheric polarity inversion line (PIL) 
pinching off first, giving rise to a progressively growing post-eruption 
loop arcade  (Kopp \& Pneumann 1976).  Chromospheric evaporation fills each 
newly reconnected loop with high-temperature plasma, which cools as 
the loop collapses; thus the hottest loops are located at the leading edge 
of the outward-expanding ``reconnection wave'' (see, e.g., Warren et al. 1999; 
Sheeley et al. 2004, 2007).

The different viewing angles afforded by the two {\it STEREO} spacecraft 
(Howard et al. 2008) provide a unique opportunity to study the 3-dimensional 
structure of coronal dimmings and post-eruption arcades.  Moreover, 
with the Extreme-Ultraviolet Imager (EUVI), these events can be observed 
simultaneously in the 17.1, 19.5, and 28.4~nm bandpasses, corresponding 
to temperatures of $\sim$1, $\sim$1.5 and $\sim$2~MK respectively, 
thereby allowing one to distinguish more easily between ``real'' dimmings 
due to mass loss (often termed transient coronal holes) and dimmings due to heating and cooling effects 
in post-eruption arcades.  We have studied several events involving 
the eruption of high-latitude filaments during the 2008--2009 
activity minimum, when the separation between the {\it STEREO}~A 
and B spacecraft was on the order of 90$^\circ$.  In this Letter, 
we focus on new results concerning the temperature dependence 
of the dimmings and the subsequent reconnection waves.

\section{EUV Observations}
\begin{figure*}\centering
\includegraphics[width=\linewidth]{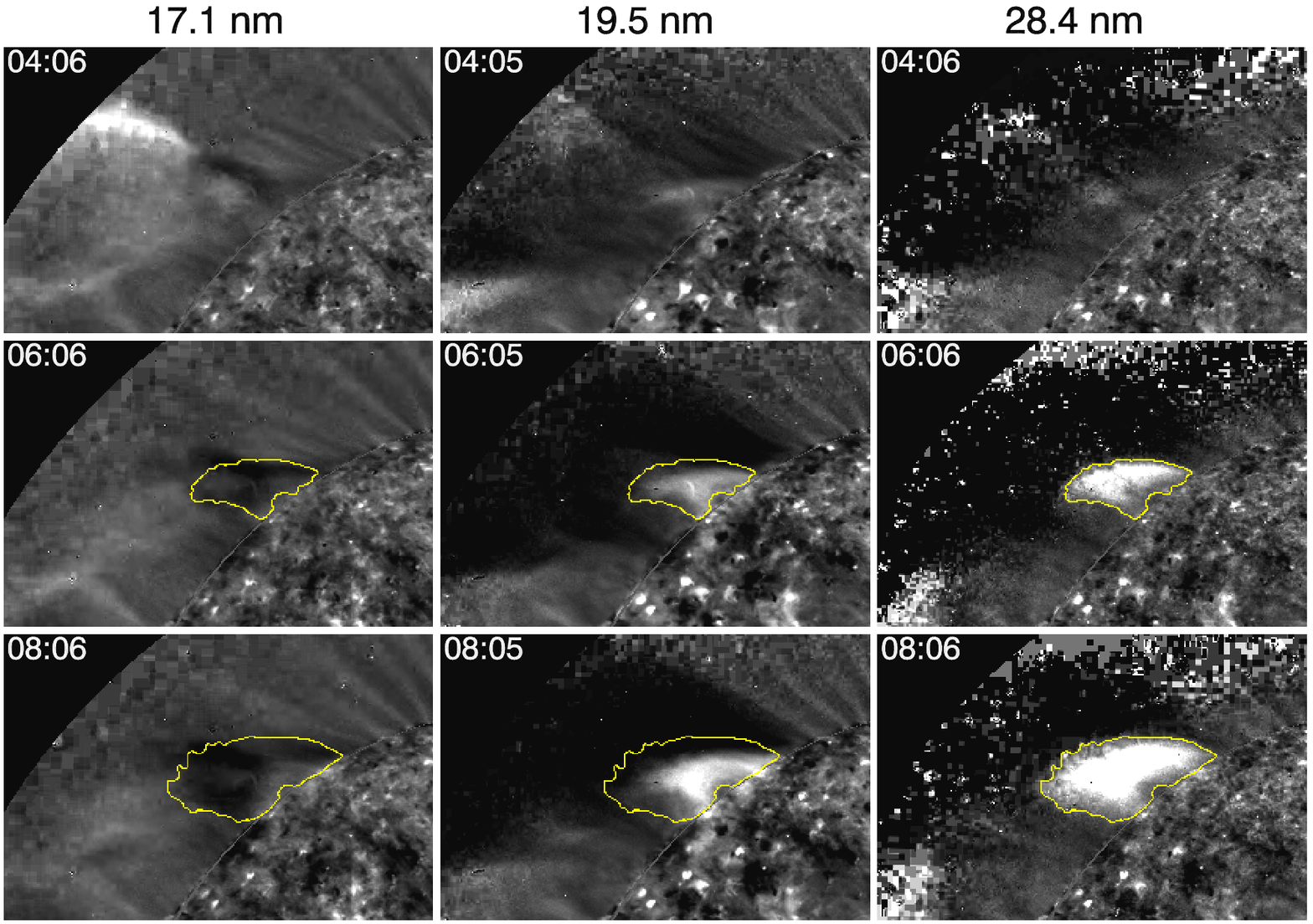}
\caption{Eruption of a filament at the northeast limb on 2009 January~14, 
as viewed from {\it STEREO}/EUVI~A.  The images show the ratio of the 
local brightness at the indicated time relative to a base image taken at 
19:06~UT  on January~13, for each of the emission lines \ion{Fe}{9} 17.1~nm 
(left column), \ion{Fe}{12} 19.5~nm (middle column), and \ion{Fe}{15} 28.4~nm 
(right column).  The on-disk areas of these base-ratio images have been 
corrected for solar differential rotation. In the two hotter lines 
(19.5 and 28.4~nm), the dimming appears as a dark void above the limb; 
this void is not seen in the cooler 17.1 nm bandpass.  The 17.1~nm images 
at 06:06 and 08:06~UT show dark features that coincide with the outer edges 
(indicated by the yellow contours) of the bright post-eruptive loops 
in 28.4~nm. [A movie is available online.] \label{fig:Jan14STA}}
\end{figure*}

The sequence of images in Figure~\ref{fig:Jan14STA} shows a filament eruption at the 
northeast limb on 2009 January~14, as observed by EUVI~A in the 
17.1, 19.5, and 28.4~nm bandpasses (see also the accompanying online movies).  Each image represents the ratio 
of the local brightness at the given time to that at 19:06~UT on 
January 13, before the start of the eruption; a shift has been applied 
to remove the effect of the photospheric differential rotation on the 
disk (Howard, Harvey and Forgach, 1990).  By taking the ratio of intensities rather than subtracting them, 
we are able to bring out features above the limb which, because of 
the rapid falloff of the coronal density with height, would not be visible 
in ordinary base-difference images.

At 04:06~UT, the last of the prominence material (which is best 
seen in \ion{He}{2} 30.4~nm) is being ejected from the limb.  This dense, 
cool material is only faintly visible at 19.5 and 28.4~nm, but appears 
in \ion{Fe}{9} 17.1~nm as a bright blob with a narrow, dark tail.  
We interpret this dark tail as a density depletion 
associated with the pinching-off of the magnetic field behind the ejection.  
Poleward and equatorward of the disconnection region, however, the corona 
is noticeably darker in the higher-temperature lines than in 17.1~nm.

At 06:06 UT (second row of images in Figure~\ref{fig:Jan14STA}), a cusp-shaped 
post-eruption arcade has begun to form in 19.5 and 28.4~nm, with the 
structure being brighter and more extended in the latter wavelength.  
In both cases, the surrounding corona (off-limb) has been strongly depleted of material 
at the given temperature, as indicated by the dark voids in the images.  
In contrast, the 17.1~nm image is dominated by neutral gray, and shows 
neither a bright arcade nor a large region of depleted density.  Instead, 
we continue to see a remnant of the wakelike depletion and some dark areas 
which lie inside the outer boundary of the bright 28.4 arcade (indicated by the yellow contours), 
and which evidently represent plasma that has been heated to temperatures 
well above 1~MK.  The same situation continues to hold as the hot 
post-eruption arcades expand (bottom row of images). 

\begin{figure*}\centering
\includegraphics[width=\linewidth]{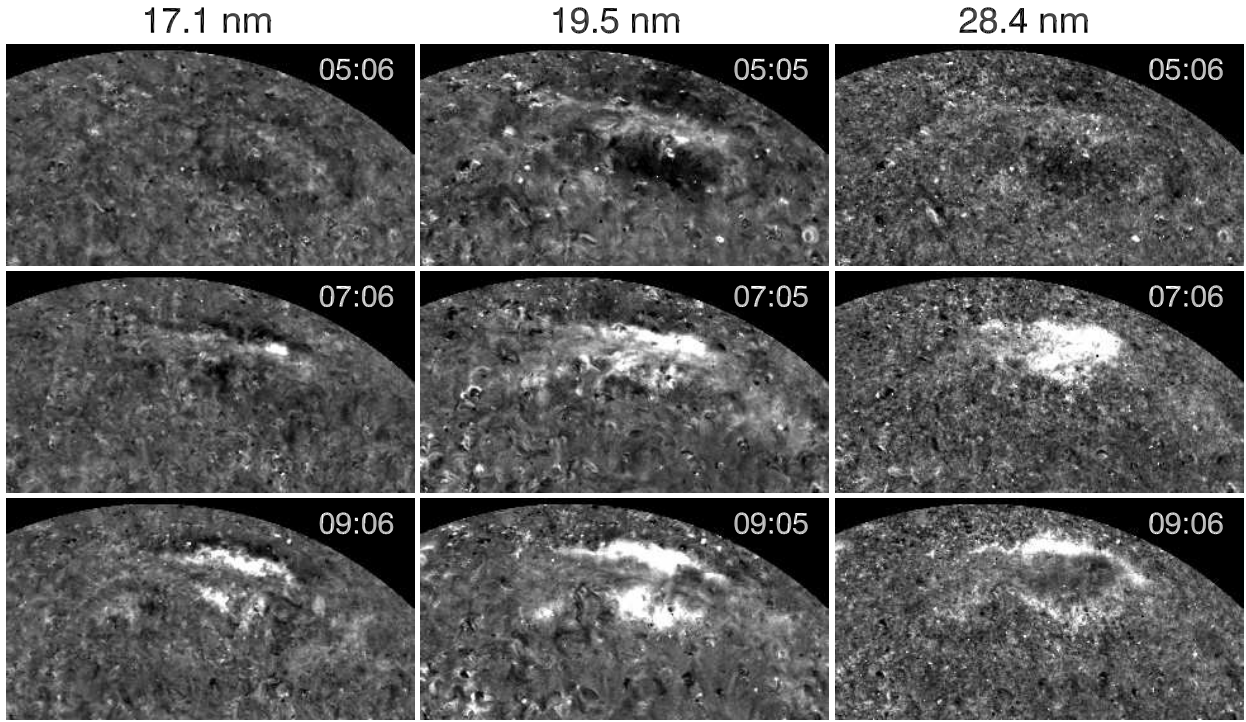}
\caption{The filament eruption of 2009 January 14, as viewed on the disk 
from {\it STEREO}/EUVI~B (compare the limb view of Figure~1).  The images 
show the ratio of the local brightness at the indicated time relative to 
that recorded 2~hr earlier, after removing the effect of solar rotation.  
The ``real'' dimming is best observed in \ion{Fe}{12} 19.5~nm at 05:05~UT; 
it is less pronounced in the 28.4~nm bandpass, which contains contributions 
from cooler lines like \ion{Si}{7} and {Mg}{7}, and is barely visible in 
\ion{Fe}{9} 17.1~nm.  At 07:06 and 09:06~UT, the 17.1~nm running-ratio images 
show a dark ``heat wave'' which coincides with the brightenings in 
28.4 and 19.5~nm.  [A movie is available online.] \label{fig:Jan14STB}}
\end{figure*}

Figure~\ref{fig:Jan14STB} shows the filament eruption and its aftermath as viewed on the 
disk from the  {\it STEREO}~B spacecraft.  Here, each image represents the 
ratio of the local brightness at the given time to that recorded 
$\sim$2~hr earlier (rather than before the eruption).  The 
\ion{Fe}{12} 19.5~nm image taken at 05:05~UT shows a pair of large, dark 
transient coronal holes, one on each side of the footpoint brightenings 
of the post-eruption arcade.  It should be noted that the brightenings 
appear in the vicinity of the PIL well before the holes (which remain essentially 
stationary) reach their darkest level.  The transient holes and 
footpoint brightenings are barely visible in the \ion{Fe}{9} 17.1~nm image, 
as expected from the corresponding limb views of Figure~\ref{fig:Jan14STA}, where the 
17.1~nm intensities undergo relatively little change during the event.  
More puzzling, at first sight, is the rather weak signature of the 
transient holes at 28.4~nm, despite the very strong darkenings that are 
seen above the limb in Figure~\ref{fig:Jan14STA}.  The weakness of the on-disk holes 
can be attributed to contributions to the 28.4~nm bandpass from 
low-temperature lines such as \ion{Si}{7} 27.5~nm, which become particularly 
significant in darker regions of the disk (see e.g., Figure~21 in 
Del~Zanna et al. 2003).

Between 05:05 and 07:05 UT, the brightenings increase in intensity 
and continue to spread through the area occupied by the transient holes. In the 17.1~nm 
running-ratio image at 07:06~UT, the poleward and (to a lesser extent) the equatorward sides 
of the footpoint brightenings are bordered by dark ribbons, which lie 
inside the boundaries of the bright 28.4 and 19.5~nm emission.  These dark 
ribbons evidently represent the footpoint areas of the 
newly reconnected loops which earlier contained \ion{Fe}{9} plasma 
but have now been heated to higher temperatures (see Fig.~\ref{fig:Jan14STA}).  
Thus the dark, outward-propagating dark ribbons observed in the relatively 
cool 17.1~nm line are a heating effect due to the closing-down of flux, 
and not a density-depletion effect due to the opening-up of 
field lines. Note that, in running-ratio images, transient holes no longer appear 
as dark areas after they have reached their darkest level in base-ratio 
images; comparing the two types of images thus offers a way to 
distinguish the ``heat waves'' from the transient coronal holes.

At 09:05 UT, the footpoint brightenings continue to propagate outward, 
but no further intensity increases occur between the diverging fronts.  
A dark ribbon is still present along the limbward side of the 
poleward-propagating 17.1~nm brightening.  Comparing the positions 
of the 19.5 and 28.4~nm brightenings, we see that the higher-latitude 
fronts appear to be shifted more relative to each other than the 
equatorward-propagating fronts.  This difference may be a projection effect 
caused by the fact that the 28.4~nm emission extends to greater heights 
than the 19.5~nm emission, and thus seems to extend farther toward 
the limb. 

\begin{figure*}\centering
\includegraphics[width=\linewidth]{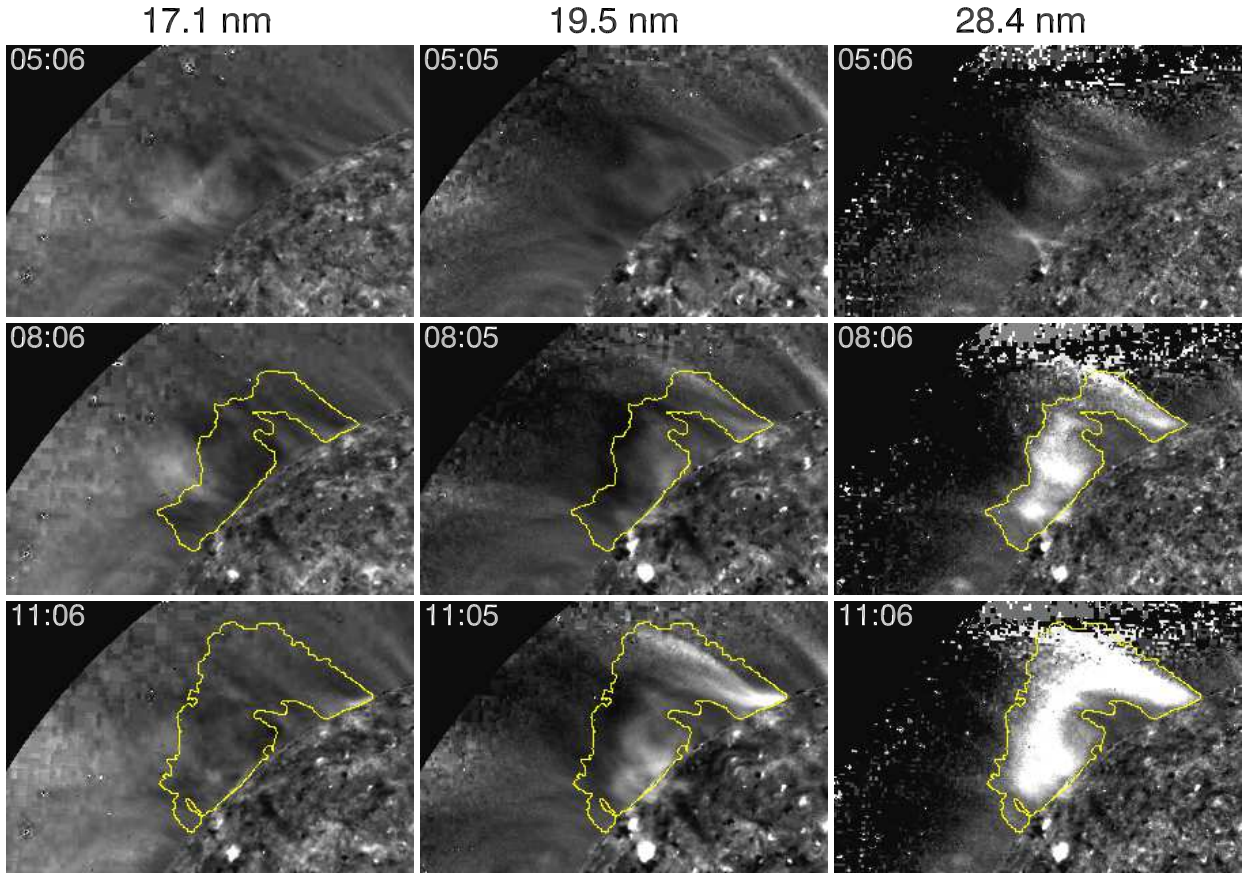}
\caption{Filament eruption at the northeast limb on 2008 December~27, 
as viewed from EUVI~ A.  The images show the ratio of local brightness
at the indicated time relative to a base image taken at 23:06~UT on 
December~26, after removing the effect of solar differential rotation from the 
on-disk area. Yellow contours indicate the boundaries of the bright 28.4~nm 
post-eruption arcade. [A movie is available online.] \label{fig:Dec27STA}}
\end{figure*}
\begin{figure*}\centering
\includegraphics[width=\linewidth]{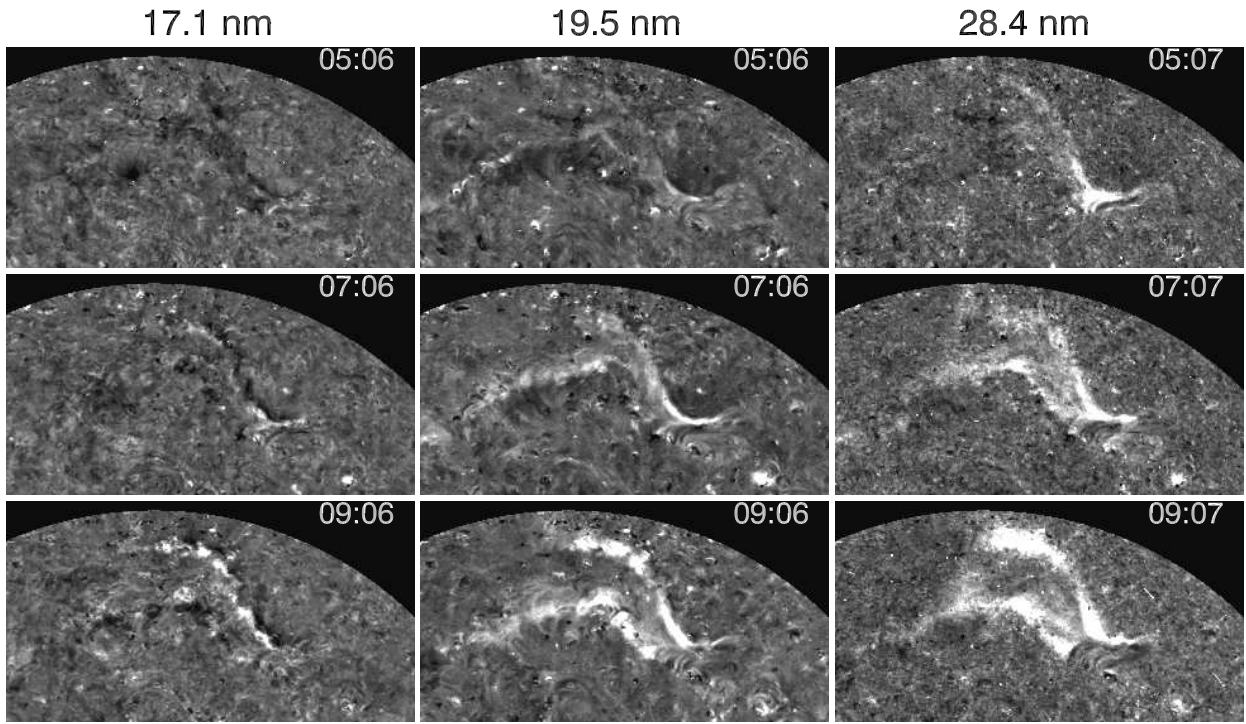}
\caption{The filament eruption of 2008 December 27, as viewed on the disk 
from EUVI~B (compare the limb view of Figure~3).  The images show the ratio 
of the local brightness at the indicated time relative to that recorded 
2~hr earlier, after correcting for solar rotation.  The ``real'' dimming 
is easiest to see in the 19.5~nm bandpass.  The 17.1~nm running-ratio images 
show a dark, outward-propagating ``heat wave''.  
[A movie is available online.] \label{fig:Dec27STB}}
\end{figure*}

As another characteristic example, Figures~\ref{fig:Dec27STA} and \ref{fig:Dec27STB} show the off-limb and 
on-disk views of a filament eruption that occurred on 2008 December~27 (see also the online movies).  
The temperature/heating effect is clearly seen in the limb view from EUVI~A 
(Figure~\ref{fig:Dec27STA}), where the post-eruption arcade is bright in 28.4~nm, fainter 
in 19.5~nm, and dark in 17.1~nm.  Conversely, the region of strongly depleted 
28.4~nm emission above the arcade appears as neutral gray in the 17.1~nm 
base-ratio images, indicating that it is mainly the hotter plasma that 
escapes when the magnetic field opens up.  Considering now the on-disk view 
from EUVI~B (Figure~\ref{fig:Dec27STB}), we first note that the dimmings in the 19.5~nm 
running-ratio images at 05:06 and 07:06~UT represent transient coronal holes.  
The faint Y-shaped darkening seen at 05:06~UT in 17.1~nm is a heating effect, 
since it coincides with a similarly shaped brightening in 28.4~nm.  
Subsequently, as this plasma cools, the double-ribbon brightening begins 
to appear in 17.1~nm, bordered on its poleward and equatorward sides by 
dark patches that represent the outward-propagating 17.1~nm ``heat wave''.

The high-latitude filament eruptions of December 27 and January 14 
both gave rise to slow CMEs observed with the white-light coronagraphs 
on  {\it STEREO}.

\section{Physical Interpretation}
\begin{figure*}\centering
\includegraphics[width=\linewidth]{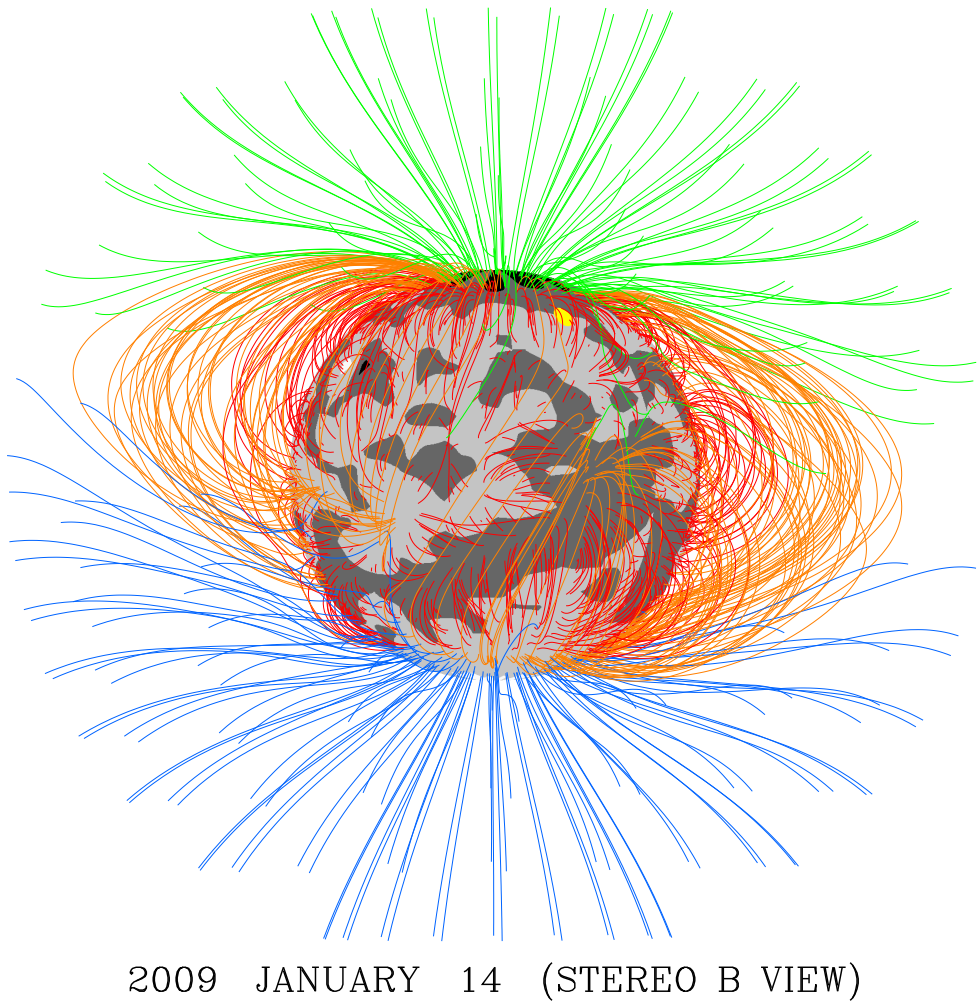}
\caption{Coronal magnetic field on 2009 January 14, as viewed from 
{\it STEREO}~B.  The field lines were derived from a PFSS extrapolation
of the MWO photospheric measurements for Carrington rotation 2079; 
the source surface was taken to be at a heliocentric distance of 
$r = 2.5$~$R_\odot$.  The yellow dot marks the approximate location 
of the filament eruption in Figure~2.  Closed loops are coded orange 
if they extend beyond $r = 1.5$~$R_\odot$, red otherwise; open field lines 
are blue (green) if they have positive (negative) polarity.  Black, 
dark gray, light gray, and white denote areas of the photosphere where 
the radial field component lies in the ranges $B_r < -6$~G, 
$-6$~G~$< B_r < 0$~G, $0$~G~$< B_r < +6$~G, and $B_r > +6$~G, respectively. \label{fig:pfss}}
\end{figure*}

Figure~\ref{fig:pfss} shows the coronal magnetic field on 2009 January 14, as viewed 
from  {\it STEREO}~B; the field lines were derived from a potential-field 
source-surface (PFSS) extrapolation of magnetograph measurements from 
the Mount Wilson Observatory (MWO).  As indicated by the yellow dot, 
the filament eruption occurred along the PIL encircling the negative-polarity 
north polar cap.  In order to account for the pair of transient coronal holes, 
most of the overlying coronal loops rooted between the polar hole boundary 
and the field-line ``part'' or separatrix on the equatorward side of the PIL must have 
opened up during the eruption.  From the fact that the post-eruption arcade 
started to form well before the twin dimmings reached their maximum strength, 
we deduce that the innermost loops near the PIL pinched off while the 
mass loss was still at an early stage. This can be seen from the off-limb 
19.5~nm ratio images in Figure~\ref{fig:Jan14STA}. At 04:05 UT the pinch-off has already occurred, while the 
evacuation of mass is still ongoing as evidenced by the darker void at 06:05 UT.

It is often assumed that all of the entrained coronal material is expelled 
when the magnetic field opens up.  That the background 17.1~nm emission 
undergoes relatively little change during filament eruptions provides 
strong evidence to the contrary: most of the cooler coronal plasma does not 
escape before the field closes down again but instead remains trapped and never leaves the Sun. This result is consistent 
with multi-wavelength Doppler observations of a transient coronal hole 
by Imada et al. (2007; see also Jin et al. 2009), using the EUV Imaging Spectrometer on {\it Hinode}.  
They found that the outflow speeds at the hole boundary were strongly 
temperature-dependent, with a steep transition from slow to fast flows 
occurring near 1~MK.  As indicated by their Figure~6, the velocity $v$ 
varies from $\sim 30$~km~s$^{-1}$ in \ion{Fe}{9} to $\sim 90$~km~s$^{-1}$ 
in \ion{Fe}{12} to $\sim 160$~km~s$^{-1}$ in \ion{Fe}{15}.  Correspondingly, 
the timescale $\tau_{\rm esc}\sim R_\odot/v$ for the plasma to travel 
a solar radius varies from $\sim$7~hr in \ion{Fe}{9} to $\sim$2~hr in  4 hr
\ion{Fe}{12} to only $\sim$1~hr in \ion{Fe}{15}.  Since the transient holes 
in the January~14 and December~27 events reached their greatest dimming 
level $\sim 3$ to $4$~hr after they first appeared in the base-ratio images, 
we conclude that the \ion{Fe}{9} emitting plasma did not have sufficient 
time to escape before the onset of reconnection.

An interesting analogy may be drawn between the temperature dependence 
of transient holes and the relationship between polar plumes and the 
interplume regions in coronal holes.  EUV plumes, which are best observed 
in lower-temperature lines like \ion{Fe}{9} 17.1~nm, are both denser and 
cooler than the interplume medium; moreover, Doppler measurements 
indicate that the flow speeds in plumes are much smaller (at low 
heights) than in the rest of the coronal hole (see e.g., Wilhelm 
et al. 1998; Cranmer et al. 1999).  Evidently, when the magnetic field 
opens up during a filament eruption, the hotter component of the corona 
behaves like the interplume medium, whereas the cooler component 
behaves like the plume gas.  According to this analogy, the hot and 
cool components would exist along separate field lines or ``strands'' 
of flux tubes.

As demonstrated by the events described in section 2, the 17.1~nm corona shows its 
most pronounced dimmings not during the eruption, but during the 
reconnection phase.  These propagating dark ribbons or ``heat waves'' occur 
at the leading edge of the 17.1~nm post-eruption arcade, and result 
from the reconnective heating of the cool plasma that was not ejected 
during the eruption.  By implication, the post-eruption brightenings 
seen at 19.5 and 28.4~nm have two different sources: reconnected loops 
that were refilled via chromospheric evaporation and underwent 
subsequent cooling, and reconnected loops that were already filled with 
cool plasma, but were then heated to temperatures well above 1~MK.

\section{Conclusions}
Our main points may be summarized as follows:

1. When viewed either on the disk or at the limb, transient coronal holes 
are much less visible in \ion{Fe}{9} 17.1~nm than in higher-temperature 
emission lines such as \ion{Fe}{12} 19.5~nm.  This observational result 
implies that most of the cooler coronal plasma does not escape 
when the magnetic field opens up. 

2. The cooler plasma remains trapped because it flows outward too slowly 
to escape before the field lines close down again.  As shown by the 
Doppler measurements of Imada et al. (2007), the outflow velocities in 
transient holes decrease dramatically at temperatures below $\sim$1~MK, 
from close to the sound speed to only $\sim$30~km~s$^{-1}$.

3. The strongest darkenings in 17.1 nm occur not during the opening-up 
of the magnetic field, but when it closes down again.  The trapped 
plasma is then heated to high temperatures, producing darkenings 
in 17.1~nm which coincide with brightenings in 28.4 and 19.5~nm.  
As the post-eruption arcade expands, a dark wavefront is observed 
in 17.1~nm at the leading edges of the arcade, which appears as an 
outward-propagating dark ribbon when viewed on the disk. Off-limb 
this gives the (wrong) impression of dark loops that are ``opening up''.

4. An analogy exists between the temperature dependence 
of transient holes and the relationship between polar plumes and the 
interplume regions in coronal holes, with the plume (interplume) behaving 
in some ways like the cool (hot) plasma in transient holes.  

Many important questions remain to be addressed.  What is the physical 
reason for the steep transition between slow and fast outflow near 
1~MK?  Is a greater fraction of the cool plasma ejected in energetic events 
involving fast CMEs?  Are dark on-disk ``heat waves'' also seen in 
higher-temperature lines such as \ion{Fe}{12} 19.5~nm, where the 
pinched-off loops are refilled by chromospheric evaporation?  
This preliminary study suggests that the continued investigation of 
temperature effects in coronal dimmings may provide a key to a 
better physical understanding of CME eruptions.

\acknowledgments We are indebted to G. A. Doschek, G. Stenborg, I. Ugarte-Urra, H. P. Warren and P. R. Young for informative discussions. The SECCHI data is produced by an international consortium of the NRL, LMSAL and NASA GSFC (USA), RAL and U. Bham (UK), MPS (Germany), CSL (Belgium), IOTA and IAS (France). This work was supported by NASA and the Office of Naval Research.


%


\end{document}